\documentstyle[12pt]{article}

\newcommand{\beg}{\begin{equation}}
\newcommand{\enq}{\end{equation}}
\newcommand{\sig}{\sigma}

\begin{document}
\title{Distribution Function for Shot Noise in Disordered Multichannel
Quantum Conductors}
\author{K. A. Muttalib$^{\dag}$ and Y. Chen$^{*}$\\
   $^{\dag}$Department of Physics, University of Florida, Gainesvile, FL
32611,
USA\\
  $^{*}$ Department of Mathematics, Imperial College, London SW7 2BZ U K}
\date{\today}
\maketitle
\begin{abstract}
We obtain the generating function for shot noise through a disordered
multi-channel conductor in the zero temperature quantum regime. The
distribution of charge transmitted over a fixed time interval is found
to be approximately Gaussian.
\end{abstract}
\newpage
\indent
The nature of shot noise in mesoscopic systems at low temperatures has
attracted considerable interest in recent years. Lesovik \cite {Lesovik}
showed that the intensity of the shot noise in a single channel two
terminal conductor is supressed in the quantum regime compared to its
classical value. This result has been generalized and applied to various
other cases \cite {Buttiker,Beenakker,Landauer}. Levitov and Lesovik
\cite {Levitov} later was able to obtain the entire distribution
function for the charge transport over a fixed time interval for a
single channel conductor and showed that the distribution function
becomes binomial in the quantum regime as opposed to Poissonian in the
classical regime. This result has been generalized to include shot noise
in a superconductor-normal metal point contact \cite {Khmelnitskii} as
well.
\par\indent
In the present work we generalize the work of ref. \cite {Levitov} and
evaluate the shot noise distribution function for a multichannel
conductor. We obtain the generating function for the distribution of
charge transmitted over a time interval $t_0$ and show that the charge
distribution is approximately Gaussian. We find that in some limit the
shot noise is reduced by a factor close to 1/3 compared to the Poisson
value as obtained in ref. \cite {Beenakker}, but we show that there is a
correction term independent of the value of conductance that can be
important in the large $t_0$ limit. Our results disagree with a recent
calculation by Lee et al \cite {Lee}, and we comment on the differences.
\par\indent
When a dc voltage $V$ is applied across the system in a two terminal
geometry, the shot noise generating function for a multichannel
conductor may be written as a product over generating function for
independent channels, each of which has a binomial distribution \cite
{Levitov}:
\beg
\chi(\lambda)=\prod_a\left(T_az+1-T_a\right)^M;\;\;\;\;\; z=e^{i\lambda},
\enq
where $M=(e/h)Vt_0$ can be interpreted as the number of attempts in time
$t_0$, and $T_a$ is the transmission probability
corresponding to channel $a$. The probability distribution for
$Q$ charges [each of charge $e$] transferred in a time $t_0$ is
\beg
P(Q,t_0)= \int_{-\pi}^{\pi}d\lambda {\rm
e}^{iQ\lambda}<\chi(\lambda,t_0)>
\enq
where the angular bracket represents an average over the distribution of
the $T_a$'s in disordered mesoscopic conductors. One can also read off
the moments or the cumulants of the charge distribution directly by
expanding $<\chi(\lambda,t_0)>$ or $\ln<\chi(\lambda,t_0)>$ in
power series of $\lambda$.
\par\indent
It has been shown \cite {Stone} that the distribution of the
transmission probability in the diffusive regime for an $N$-channel
disordered conductor  can be obtained from an appropriate random matrix
theory constructed for the matrix $X=[TT^{\dag}+(TT^{\dag})^{-1}-2I]/4$
where $T$ is the $2N\times 2N$
transfer matrix describing the conductor and $I$ is
the identity matrix.
The $N$ doubly degenerate real non-negative eigenvalues $x_a$ of the
matrix $X$ are related to the transmission probability $T_a$ by the
relation
\beg
T_a=\frac{1}{1+x_a};\;\;\;\;\; 0 \leq x_a \leq \infty.
\enq
According to the  random matrix theory \cite {Mehta}, the joint
probability distribution for the $N$ eigenvalues $x_a$ can then be
written in the form
\beg
p(x_1\cdots x_N)=\prod_{a<b}|x_a-x_b|^{\alpha}\prod_c e^{-V(x_c)},
\enq
where $V(x)$ is a Lagrange multiplier function \cite {Balian} that
incorporates physical constraints like a given value of the first
moment, and $\alpha$ is a symmetry parameter. The value of  $\alpha$ is
1,2 or 4 depending on whether the symmetry of the matrix ensemble is
orthogonal, unitary or symplectic. A reasonably good description of a
disordered conductor in the
diffusive regime \cite {Stone,Chen2,Bee}, is obtained from the choice
$V(x)=tx$,
where the  parameter $t$ is
related to the dimensionless ohmic conductance $g_0$ of the system by
$t=\alpha g_0^2/2N$. The conductance $g$ is given by the relation
$g=\sum_a \frac{1}{1+x_a},$ and is therefore a linear statistic of the
eigenvalues $x_a$; the variance of such linear statistics have well
known properties \cite {Mehta,Basor,Beans}.
\par\indent
The generating function for shot noise is, however, not a linear
statistic; it is a product statistic as given by eq. (1). To our
knowledge, fluctuation properties of such product statistics within a
random matrix framework have not been calculated before. [One might
attempt to avoid the problem by evaluating the average  $<\ln\chi>$,
which is a linear statistic \cite {Lee}, instead of the appropriate
$\ln<\chi>$; this is a reasonable approximation only if the
distribution is very sharply peaked. Our results differ
from those of \cite {Lee}]. We will evaluate the average of the product
statistic within the random matrix framework using a continuum
approach \cite {Dyson}, which is known to work very well in the
diffusive regime. In this approach the probability distribution (4) is written
as the exponential of a fictitious Hamiltonian where the eigenvalues are
considered as particles repelling each
other logarithmically as in a Coulomb fluid, within a confining potential
$V(x)$, at
a temperature $1/\alpha$.
\par\indent
In terms of the variable $x_a$, the generating function can be rewritten
as
\beg
\chi(\lambda,t_0)={\rm e}^{M\sum_a\ln[(x_a+z)/(x_a+1)]}.
\enq
The average of $\chi$ over the  distribution of $x_a$ can then be
written as
\beg
\left<\chi(\lambda)\right>=Z_M/Z_0={\rm e}^{-(F_M-F_0)}
\enq
where
\beg
Z_M=\int_{0}^{\infty}\prod_a dx_a {\rm exp}\left[\alpha \sum_{a<b}
\ln|x_a-x_b|-t\sum_a x_a+M\sum_a\ln\frac{x_a+z}{x_a+1}\right].
\enq
Here $Z_0$, defined as $Z_{M=0}$ is the partition function and $F_0$,
defined as $F_{M=0}$ is the free energy for the coulomb fluid
characterized by the logarithmic repulsion and a linear confining
potential as shown in eq. (7) for $M=0$. It is well known that for a
linear confining potential the density of eigenvalues at the origin
scales with the number of eigenvalues $N$. Therefore the large N
continuum approximation is valid \cite {Dyson}, and we can use the
coulomb fluid results
\beg
U^{\prime}_M(x)-\alpha\int_I\frac{dy}{x-y}\sig_M(y)=0,
\enq
where $\sig_M(x)$ is the `effective' density of eigenvalues contained
in an interval $I$, the prime
denotes a derivative with respect to $x$, and
\beg
U_M(x)=U_0(x)+U_c(x)=V(x)-M\ln\left[\frac{x+z}{x+1}\right]
\enq
is the `effective' confining potential, which includes the confining
potential $U_0(x)=U_{M=0}(x)=V(x),$ as well as the `correction'
$U_c(x)=M\ln\left[\frac{x+z}{x+1}\right]$. [We use the word `effective'
within a quotation mark because the parameter $z$ is complex, and the
density or the confining potential has only a mathematical meaning.] The
range of integration $I$ is determined by the allowed range of eigenvalues as
well as the normalization requirement. In
terms of the density, the free energy $F_M$ can be obtained \cite {Chen}
from the relation
\beg
F_M=(1/2)A_MN+\int_{I}dx\sig_M(x)U_M(x),
\enq
where $A_M$ is the chemical potential determined from the integration
constant,
\beg
U_M(x)-\alpha\int_Idy\sig_M(y)\ln|x-y|=A_M.
\enq
The
general solution to equation (8) for the density is given by $[I=(0,b)]$
\beg
\sig_M(x)=\frac{1}{\pi^2\alpha}{\sqrt {\frac{b-x}{x}}}
\int_{0}^{b}\frac{dy}{y-x}{\sqrt {\frac{y}{b-y}}}U_M^{\prime}(y),
\enq
where $b$ is the upper limit for the density set by the
symmetry and normalization requirements.
\par\indent
To obtain explicit expressions, we write
\beg
\sig_M(x)=\sig_0(x)+\sig_c(x),
\enq
where $\sig_0(x)=\sig_{M=0}(x)$ is the density for the potential
$U_0(x)$ alone, and $\sig_c(x)$ is the correction to the density due
to $U_c(x)$. Then for $U_0(x)=tx$, and for non-negative
eigenvalues, we obtain
\beg
\sig_0(x)=\frac{t}{\pi\alpha}{\sqrt {\frac{b-x}{x}}},
\enq
and
\beg
\sig_c(x)=\frac{M(1-{\sqrt z})}{\pi\alpha{\sqrt x}}\frac{x-{\sqrt
z}}{(x+1)(x+z)};\;\;\; |arg z|<\pi.
\enq
The normalization condition $\int_{0}^{b}dx\sig_M(x)=N$ gives us the
value for b, $N=tb/2\alpha.$ Note that the contribution from $\sig_c$ to
the normalization integral is zero\cite{remark},
\beg
\int_{0}^{\infty}dx\sig_c(x)=0.
\enq
Therefore we may conclude that the chemical potential $A_M$ differs from
$A_0$ by a
negligible amount as $N\rightarrow \infty.$ This provides a major
simplification, and
the change in free energy is then given by
\beg
F_M-F_0=(1/2)\int_{0}^{b}dx\sig_0U_c+(1/2)\int_{0}^{\infty}dx\sig_cU_0
+(1/2)\int_{0}^{\infty}dx\sig_cU_c,
\enq
where $\sig_0$, $U_0$, $\sig_c$ and $U_c$ are as defined before.
In the limit of large system size, $b=4N^2/g_0^2\rightarrow \infty$
($g_0=Nl/L$, where $l$ is the mean free path and $L$ is the system size),
all integrals can be explicitly evaluated,
and the generating function is given by
\beg
\left<\chi(\lambda)\right>={\rm exp}\left[-\frac{1+\pi}{2\pi}Mg_0(1-{\sqrt
z})+\frac{Mg_0^2}{4N}(1-z)+\frac{4M^2}{\alpha}\ln\frac{1+{\sqrt
z}}{2z^{1/4}}\right],\;\;z=e^{i\lambda},\;\;|\lambda|<\pi
\enq
where we have used the conductance $g_0$ instead of the parameter $t$.
This is our principal result. Again for large system size,
$\frac{Mg_0^2}{N}<<Mg_0,$ and the term proportional to $(1-z)$ can
be neglected compared to the term proportional to $(1-{\sqrt z})$.
However, for long enough measurement time $t_0$ (a single energy
generating function can be considered only in this limit), the term
proportional to $M^2$ may not be negligible ($M=eVt_0/h$). Note that
this term is independent of the conductance $g_0$, but depends on the symmetry
parameter $\alpha$.
\par\indent
We can use eq. (2) to evaluate an approximate probability distribution
$P(Q,t_0)$. We use the fact that major contributions to the integral in
(2) come from regions around $\lambda=0$ where expansions in powers of
$\lambda$ can be used. If we keep terms upto order $\lambda^2$, the
limits of the integral can then be extended to $\pm\infty$. The result,
valid around the mean $Q_0$, is a Gaussian distribution:
\beg
P(Q,t_0)\sim {\rm exp}
\left[-\frac{(Q-Q_0)^2}{2\left(\frac{Q_0}{2}+\frac{M^2}{8\alpha}\right)}\right]
\enq
where  the mean value
\beg
<Q>=Q_0=\frac{1+\pi}{2\pi}Mg_0.
\enq
We can also expand $\ln\left<\chi\right>$ in  power series of
$\lambda$ to obtain directly the cumulants of the charge distribution.
For $\frac{Mg_0^2}{N}\rightarrow 0$, we get
\beg
\ln\left<\chi\right>
=Q_0x+\left(\frac{Q_0}{2}+\frac{M^2}{8\alpha}\right)\frac{x^2}{2!}+
\left(\frac{Q_0}{4}\right)\frac{x^3}{3!}+
\left(\frac{Q_0}{8}-\frac{M^2}{64\alpha}\right)\frac{x^4}{4!}+\ldots
\enq
where the cumulant $<<Q^k>>$ can simply be read off as the
coefficient of $x^k/k!$ in the expansion (21). The first two cumulants
agree with the approximate Gaussian distribution obtained above. Note
that $Q_0$ defines the average charge transferred in time $t_0$, while
the intensity of the shot noise is given by $<<Q^2>>$. A
completely uncorrelated classical system would have a Poisson
distribution \cite {Beenakker} with $<Q>$ as well as $<<Q^2>>$
given by  $Q_P=Mg_0$. If the term proportional to $M^2$ is neglected,
the shot noise is reduced compared to $Q_P$ by a factor
$\frac{1+\pi}{4\pi}$, which is close to 1/3 obtained in ref. \cite
{Beenakker}. However the correction term, which is independent of $g_0$,
may become important for long enough observation time and/or stronger
disorder. On the other hand, for stronger disorder, the appropriate
random matrix model \cite {Chen2} has a confining potential $V(x)\sim
[\ln x]^2,$ for very large $x$, and the density near the origin no
longer scales with $N$.  In this regime the usual large $N$ continuum
approximation breaks down. The shot noise distribution in this regime is
under investigation.
\par\indent
We thank Selman Herschfield for many discussions. YC thanks the
Institute for Fundamental Theory at the University of Florida for its
kind hospitality and partial support during his visit.


\begin{thebibliography}{99}
\bibitem{remark}Since the integrand decreases rapidly to zero as $x$
increases,
the upper limit can be pushed to $\infty.$
\end{thebibliography}
\end{document}